\documentclass[12pt]{article}
\usepackage{amsmath}

\begin{document}

\title{\bf On integrability of the vector short pulse equation}

\author{{\sc Sergei Sakovich}\\[6pt]
{\small Institute of Physics, National Academy of Sciences, 220072 Minsk, Belarus}\\[6pt]
{\small E-mail: saks@tut.by}}

\date{}

\maketitle

\begin{abstract}
Using the Painlev\'{e} analysis preceded by appropriate transformations of nonlinear systems under investigation, we discover two new cases in which the Pietrzyk--Kanatt\v{s}ikov--Bandelow vector short pulse equation must be integrable due to the results of the Painlev\'{e} test. Those cases are technologically important because they correspond to the propagation of polarized ultra-short light pulses in usual isotropic silica optical fibers.
\end{abstract}

\bigskip \bigskip \bigskip

The short pulse equation (SPE), which has the form
\begin{equation}
u_{xt} = u + \tfrac{1}{6} \left( u^3 \right)_{xx} \label{spe}
\end{equation}
up to a scale transformation of its variables, was introduced recently by Sch\"{a}fer and Wayne~\cite{SW} as a model equation describing the propagation of ultra-short light pulses in silica optical fibers (note, however, that for the first time this equation appeared in differential geometry, as one of Rabelo's equations describing pseudospherical surfaces~\cite{Rab}). Unlike the celebrated nonlinear Schr\"{o}dinger equation which models the evolution of slowly varying wave trains, the SPE is well applicable when the pulse spectrum is not narrowly localized around the carrier frequency, that is when the pulse is as short as a few cycles of its central frequency. Such ultra-short pulses are very important for future technologies of ultra-fast optical transmission of information. The SPE~\eqref{spe} is an integrable equation possessing a Lax pair~\cite{BRT,SS1} of the Wadati--Konno--Ichikawa type~\cite{WKI}. The transformation between the SPE and the sine-Gordon equation was discovered in~\cite{SS1}, and later it was used in~\cite{SS2} for obtaining exact loop and pulse solutions of the SPE from the well-known kink and breather solutions of the sine-Gordon equation. The derivation of that transformation was considerably simplified in~\cite{SS3}, where analogous transformations to well-studied equations were also found for the other three equations of Rabelo. The recursion operator~\cite{SS1}, Hamiltonian structures and conserved quantities~\cite{B1,B2}, Hirota's bilinear representation~\cite{KBK}, multisoliton solutions~\cite{M1} and periodic solutions~\cite{Par,M2} of the SPE~\eqref{spe} were found and studied as well.

Very recently, Pietrzyk, Kanatt\v{s}ikov and Bandelow~\cite{PKB} introduced the vector short pulse equation (VSPE), a two-component nonlinear wave equation that generalizes the scalar SPE~\eqref{spe} and describes the propagation of polarized ultra-short light pulses in cubically nonlinear anisotropic optical fibers, which can be written as
\begin{equation}
U_{n,xt} = c_{ni} U_i + c_{nijk} \left( U_i U_j U_k \right)_{xx} , \label{vspe}
\end{equation}
where $n,i,j,k = 1,2$, and the summation over the repeated indices is assumed. Since the constant coefficients $c_{ni}$ and $c_{nijk}$ are determined by optical properties of the fiber's material, there is a wide variety of mathematically different cases of the VSPE~\eqref{vspe}, and it is interesting to find out which of them are integrable systems of coupled nonlinear wave equations. Three integrable cases of the VSPE~\eqref{vspe} were obtained in~\cite{PKB} by direct construction of their Lax pairs, namely,
\begin{equation}
u_{xt} = u + \tfrac{1}{6} \left( u^3 + 3 u v^2 \right)_{xx} , \qquad v_{xt} = v + \tfrac{1}{6} \left( 3 u^2 v + v^3 \right)_{xx} , \label{vspe1}
\end{equation}
\begin{equation}
u_{xt} = u + \tfrac{1}{6} \left( u^3 - 3 u v^2 \right)_{xx} , \qquad v_{xt} = v + \tfrac{1}{6} \left( 3 u^2 v - v^3 \right)_{xx} , \label{vspe2}
\end{equation}
and
\begin{equation}
u_{xt} = u + \tfrac{1}{6} \left( u^3 \right)_{xx} , \qquad v_{xt} = v + \tfrac{1}{2} \left( u^2 v \right)_{xx} , \label{vspe3}
\end{equation}
where $u$ and $v$ denote the polarization components $U_1$ and $U_2$. The following valuable remark was made in~\cite{PKB} on the nature of the system~\eqref{vspe3}: this case of the VSPE describes the propagation of a small perturbation $v$ on the background of a solution $u$ of the scalar SPE~\eqref{spe}. Contrary to what was proposed in~\cite{PKB}, however, we cannot consider the VSPE~\eqref{vspe1} as a short pulse counterpart of the Manakov system~\cite{Man} of coupled nonlinear Schr\"{o}dinger equations. Indeed, in the new variables $p = u + v$ and $q = u - v$ the equations of the system~\eqref{vspe1} become uncoupled and turn into two scalar SPEs~\eqref{spe} for $p$ and $q$ separately, whereas the polarization modes in the Manakov system do interact nonlinearly~\cite{APT}. In the variables $p = u + i v$ and $q = u - i v$ the equations of the system~\eqref{vspe2} become uncoupled as well. It is easy to prove that the $4 \times 4$ zero-curvature representations, found in~\cite{PKB} for~\eqref{vspe1} and~\eqref{vspe2}, can be brought by gauge transformations into the block-diagonal form with the $2 \times 2$ diagonal blocks corresponding to the zero-curvature representation of the scalar SPE~\eqref{spe} for $p$ or $q$.

In the present paper, we show that there are at least two more cases of the VSPE~\eqref{vspe} which can be strongly expected to be integrable due to the analytic properties of their general solutions, namely,
\begin{equation}
u_{xt} = u + \tfrac{1}{6} \left( u^3 + u v^2 \right)_{xx} , \qquad v_{xt} = v + \tfrac{1}{6} \left( u^2 v + v^3 \right)_{xx} , \label{my1}
\end{equation}
and
\begin{equation}
u_{xt} = u + \tfrac{1}{6} \left( u^3 \right)_{xx} , \qquad v_{xt} = v + \tfrac{1}{6} \left( u^2 v \right)_{xx} . \label{my2}
\end{equation}
The VSPE~\eqref{my1} represents the technologically important case, where the fiber is made of a cubically nonlinear isotropic optical material, such as the widely used silica glass, but the ultra-short light pulse is not linearly polarized; if integrable, this system can be interesting as a short pulse counterpart of the Manakov system. The VSPE~\eqref{my2} can be considered as the limiting case of the system~\eqref{my1} for small values of $v$, that is the case of almost linearly polarized pulses. We discover these systems~\eqref{my1} and~\eqref{my2} by applying the Painlev\'{e} test for integrability of partial differential equations~\cite{WTC,Tab,RGB} to the following two one-parameter classes of VSPEs:
\begin{equation}
u_{xt} = u + \tfrac{1}{6} \left( u^3 + c u v^2 \right)_{xx} , \qquad v_{xt} = v + \tfrac{1}{6} \left( c u^2 v + v^3 \right)_{xx} , \label{class1}
\end{equation}
and
\begin{equation}
u_{xt} = u + \tfrac{1}{6} \left( u^3 \right)_{xx} , \qquad v_{xt} = v + \tfrac{c}{6} \left( u^2 v \right)_{xx} , \label{class2}
\end{equation}
where $c$ is the parameter. Of course, this is far not the complete test for integrability of the whole variety of systems~\eqref{vspe} but rather the first attempt to search for new integrable VSPEs systematically. In what follows, we make the computations using the Mathematica computer algebra system~\cite{Wol} and omit their inessential bulky details.

Let us consider the class of systems~\eqref{class1} first. It is easy to see that the Painlev\'{e} test cannot be applied to the VSPE~\eqref{class1} directly, for the reason of an inappropriate dominant behavior of solutions near a movable singularity manifold, and we must appropriately transform the nonlinear system under investigation in order to improve this behavior and start the test. This is a point of crucial importance in our study. We follow the way of transformation similar to the way used in~\cite{SS3} for the scalar SPE~\eqref{spe} and other Rabelo's equations. We make the change of the independent variable $x$,
\begin{equation}
u(x,t) = f(y,t) , \qquad v(x,t) = g(y,t), \qquad y = y(x,t) , \label{chx}
\end{equation}
and determine the function $y(x,t)$ by the relation
\begin{equation}
y_t = \tfrac{1}{2} \left( u^2 + v^2 \right) y_x . \label{y1}
\end{equation}
In this relation~\eqref{y1}, the polynomial $u^2 + v^2$ is taken for the reason of symmetry between $u$ and $v$; one could use there a general quadratic polynomial in $u$ and $v$ instead, but this would have no effect on the dominant behavior of solutions, positions of resonances and compatibility of recursion relations, found during the Painlev\'{e} analysis. Then, inverting $y = y(x,t)$ as $x = x(y,t)$, we obtain from~\eqref{class1} and~\eqref{y1} the following system of three coupled equations for $f(y,t)$, $g(y,t)$ and $x(y,t)$:
\begin{gather}
2 x_t + f^2 + g^2 = 0 , \notag \\[3pt]
6 x_y^2 f_{yt} + (3-c) g^2 x_y f_{yy} - 2c f g x_y g_{yy} + \left( (c-3) g^2 f_y + 2c f g g_y \right) x_{yy} \notag \\
+ (6-4c) g f_y g_y x_y - 2c f g_y^2 x_y - 6 f x_y^3 = 0 , \notag \\[3pt]
6 x_y^2 g_{yt} + (3-c) f^2 x_y g_{yy} - 2c g f x_y f_{yy} + \left( (c-3) f^2 g_y + 2c g f f_y \right) x_{yy} \notag \\
+ (6-4c) f g_y f_y x_y - 2c g f_y^2 x_y - 6 g x_y^3 = 0 . \label{trc1}
\end{gather}
Note that the fact of correspondence between the fifth-order system~\eqref{trc1} and the fourth-order system~\eqref{class1} (we mean here the total order of a system, or the number of arbitrary functions in its general solution) can be explained by the invariance of~\eqref{trc1} with respect to an arbitrary transformation $y \mapsto Y(y)$, which just means that the solutions of~\eqref{trc1} represent the solutions of~\eqref{class1} parametrically, with $y$ being the parameter.

Substitution of the expansions
\begin{align}
x & = x_0 (y,t) \phi(y,t)^{\alpha} + \dotsb + x_r (y,t) \phi(y,t)^{r+\alpha} + \dotsb , \notag \\
f & = f_0 (y,t) \phi(y,t)^{\beta} + \dotsb + f_r (y,t) \phi(y,t)^{r+\beta} + \dotsb , \notag \\
g & = g_0 (y,t) \phi(y,t)^{\gamma} + \dotsb + g_r (y,t) \phi(y,t)^{r+\gamma} + \dotsb \label{sexp}
\end{align}
to the system~\eqref{trc1} determines the dominant behavior of solutions in the neighborhood of a manifold $\phi (y,t) = 0$, i.e.\ the admissible values of $\alpha$, $\beta$, $\gamma$, $x_0$, $f_0$ and $g_0$, and the corresponding positions of resonances $r$, where some arbitrary functions can enter the expansions. If we assume that at least one of the exponents $\alpha$, $\beta$, $\gamma$ is negative, we immediately get $\alpha = \beta = \gamma = -1$ for all values of $c$ except $c = -1$ (we do not know at present how to transform the systems~\eqref{class1} and \eqref{class2} with $c = -1$ in order to start the Painlev\'{e} test for them, and we do not consider the case of $c = -1$ in this paper). We have to exclude the manifolds $\phi = 0$ with $\phi_y \phi_t = 0$, for which no well-posed recursion relations appear for the coefficients $x_n$, $f_n$, $g_n$ of the expansions~\eqref{sexp}; the reason to exclude such characteristic manifolds consists in that arbitrarily nasty singularities of solutions can occur along characteristics, even for integrable equations~\cite{War,S1}. For non-characteristic manifolds, without loss of generality, we choose $\phi_y (y,t) = 1$ with $\phi_t \neq 0$ and set all the coefficients $x_n$, $f_n$, $g_n$ in~\eqref{sexp} to be functions of $t$ only. Then we find for $c \neq -1, 1, 3$ that
\begin{gather}
x_0 = - (1+c) \phi_t , \qquad f_0 = \pm {\mathrm i} \sqrt{1+c} \, \phi_t , \qquad g_0 = \pm {\mathrm i} \sqrt{1+c} \, \phi_t , \notag \\
r = -1 , 1 , 4 , \tfrac{1}{2} \left( 5 - \sqrt{\tfrac{27+23c}{3-c}} \right) , \tfrac{1}{2} \left( 5 + \sqrt{\tfrac{27+23c}{3-c}} \right) , \label{symc}
\end{gather}
where the $\pm$ signs in the expressions for $f_0$ and $g_0$ are independent, and ${\mathrm i}^2 = -1$. The case of $c = 3$, i.e.\ the VSPE~\eqref{vspe1}, is not of interest because the equations can be easily uncoupled. And for $c = 1$ we find that
\begin{equation}
x_0 = -2 \phi_t , \qquad f_0^2 + g_0^2 = -4 \phi_t^2 , \qquad r = -1 , 0, 1 , 4 , 5 , \label{sym1}
\end{equation}
where $g_0 (t)$ (or $f_0 (t)$) is arbitrary due to the resonance $r=0$. In the case of relations~\eqref{symc}, there are only three possibilities to have four resonances in integer non-negative positions, namely, $c = -1, 0, 1$. However, the values $c = -1, 1$ have been excluded, whereas the value $c = 0$ corresponds to the uninteresting case of uncoupled equations in the VSPE~\eqref{class1}. On the other hand, in the case of relations~\eqref{sym1} corresponding to $c = 1$, we have already got the admissible positions of resonances. Continuing to study this case, we derive from the equations~\eqref{trc1} with $c = 1$ the recursion relations for the coefficients of expansions~\eqref{sexp} with $\alpha = \beta = \gamma = -1$, check the compatibility of those relations at the resonances, and find that all the compatibility conditions are satisfied identically. Thus, in this case, the singular expansions of solutions turn out to be some generalized Laurent series containing no non-integer powers and no logarithmic terms, the arbitrary functions in the series being $g_0 (t)$, $x_1 (t)$, $g_4 (t)$, $g_5 (t)$, and $\psi (t)$ in $\phi = y + \psi (t)$.

To complete the Painlev\'{e} test for the system~\eqref{trc1} with $c = 1$, we also have to study the expansions~\eqref{sexp} with $\alpha = \beta = \gamma = 0$. Such expansions, which start like Taylor series, exist for every partial differential equation. Usually, they are the Taylor expansions of regular solutions governed by the Cauchy--Kovalevskaya theorem~\cite{Pet}. In some cases, however, such Taylor-like expansions can contain non-dominant singularities; this may happen, for example, when the Kovalevskaya form of a studied equation is singular for some of Cauchy data~\cite{Cla,DKLN}. Substituting the expansions~\eqref{sexp} with $\alpha = \beta = \gamma = 0$ to the system~\eqref{trc1} with $c = 1$, we find that no well-posed recursion relations appear for the coefficients $x_n$, $f_n$, $g_n$ if the manifold $\phi = 0$ is determined by the conditions $\phi_y \phi_t = 0$ or $\left( f_0^2 + g_0^2 + 6 x_{0,t} \right) \phi_y = 6 x_{0,y} \phi_t$. We exclude such characteristic manifolds, choosing $\phi_y = 1$ with $\phi_t \neq 0$ and setting all the coefficients $x_n$, $f_n$, $g_n$ in~\eqref{sexp} to be functions of $t$ only, with $f_0^2 + g_0^2 + 6 x_{0,t} \neq 0$. Then we find that the positions of resonances depend on whether the Cauchy data satisfy the condition $f_0^2 + g_0^2 + 2 x_{0,t} = 0$. When $f_0^2 + g_0^2 \neq - 2 x_{0,t}$, the arbitrary functions in the expansions are $x_0 (t)$, $f_0 (t)$, $g_0 (t)$, $f_1 (t)$, $g_1 (t)$, as well as $\psi (t)$ in $\phi = y + \psi (t)$ (the appearance of one extra arbitrary function in Taylor-like expansions of solutions was discussed in~\cite{Cla}). When $f_0^2 + g_0^2 = - 2 x_{0,t}$, the arbitrary functions in the expansions are $x_0 (t)$, $g_0 (t)$, $g_1 (t)$, $g_2 (t)$, as well as $\psi (t)$ in $\phi = y + \psi (t)$. In both cases, the positions of resonances are integer, and the recursion relations are compatible at the resonances. Therefore the expansions are some Taylor series containing no non-integer powers and no logarithmic terms. Consequently, the system~\eqref{trc1} with $c = 1$ has passed the Painlev\'{e} test, and the VSPE~\eqref{my1} can be strongly expected to be integrable.

Let us proceed now to the class of systems~\eqref{class2}. Making the same transformation~\eqref{chx}, setting
\begin{equation}
y_t = \tfrac{1}{2} u^2 y_x , \label{y2}
\end{equation}
and inverting $y = y(x,t)$ as $x = x(y,t)$, we obtain from~\eqref{class2} and~\eqref{y2} the following system of three coupled equations for $f(y,t)$, $g(y,t)$ and $x(y,t)$:
\begin{gather}
2 x_t + f^2 = 0 , \qquad f_{yt} - f x_y = 0 , \notag \\[3pt]
6 x_y^2 g_{yt} + (3-c) f^2 x_y g_{yy} - 2c g f x_y f_{yy} + \left( (c-3) f^2 g_y + 2c g f f_y \right) x_{yy} \notag \\
+ (6-4c) f g_y f_y x_y - 2c g f_y^2 x_y - 6 g x_y^3 = 0 . \label{trc2}
\end{gather}
Then, using the expansions~\eqref{sexp}, choosing $\phi_y = 1$ with $\phi_t \neq 0$ to exclude characteristic manifolds $\phi = 0$, setting all the coefficients $x_n$, $f_n$, $g_n$ to be functions of $t$ only, and assuming that at least one of the exponents $\alpha$, $\beta$, $\gamma$ is negative, we find from the system~\eqref{trc2} that
\begin{gather}
\alpha = \beta = -1 , \qquad \gamma^2 - 3 \gamma + 2 = 6 / c , \notag \\
x_0 = - 2 \phi_t , \qquad f_0 = \pm 2 {\mathrm i} \phi_t , \qquad r = -1 , 0 , 1 , 4 , 3 - 2 \gamma , \label{asym}
\end{gather}
where the resonance $r = 0$ corresponds to the arbitrariness of $g_0 (t)$. The resonance $r = 3 - 2 \gamma$ must correspond to the arbitrary coefficient $g_{3 - 2 \gamma} (t)$, due to the structure of recursion relations which follow from~\eqref{trc2} and~\eqref{sexp}. Denoting the resonance position $3 - 2 \gamma$ as $m$ and using~\eqref{asym}, we have
\begin{equation}
\gamma = ( 3 - m ) / 2 , \qquad c = 24 / ( m^2 - 1 ) . \label{gcm}
\end{equation}
The admissible values of $m$ are $m = 2, 3, 4, 5, \dotsc$, since $m=0$ is excluded because it leads to the double resonance $r = 0, 0$ which indicates that the expansion for $g$ must contain a logarithmic term, and $m=1$ is excluded because it implies $c = \infty$. Though the even values of $m$ correspond to non-integer values of $\gamma$, these cases should not be excluded, because by introducing the new variable $h = g^2$ one can improve the dominant behavior of solutions.

We have found infinitely many cases of the system~\eqref{trc2}, which are all characterized by some admissible dominant behavior of solutions and admissible positions of resonances. Unfortunately, we cannot check the compatibility of the recursion relations at the resonances for the whole infinite set of those cases at once. This situation is quite similar to the one observed in the Painlev\'{e} analysis of triangular systems of coupled Korteweg--de~Vries equations~\cite{S2}. On available computers, we were able to complete the Painlev\'{e} test for the cases $m = 2, 3, \dotsc , 9, 10$ of the system~\eqref{trc2} with $c$ given by~\eqref{gcm}. The recursion relations turn out to be compatible only in the cases $m=3$ and $m=5$, whereas some nontrivial compatibility conditions appear in all other cases at the resonance $r=m$ as an indication of non-dominant logarithmic singularities of solutions. The case $m=3$ with $\gamma = 0$ and $c=3$ corresponds to the integrable VSPE~\eqref{vspe3} discovered in~\cite{PKB}. The case $m=5$ with $\gamma = -1$ and $c=1$ corresponds to our new VSPE~\eqref{my2}. Continuing to study the system~\eqref{trc2} with $c=1$, we consider the expansions~\eqref{sexp} with $\alpha = \beta = \gamma = 0$, and they turn out to be some Taylor series containing no non-integer powers and no logarithmic terms. Consequently, the system~\eqref{trc2} with $c = 1$ has passed the Painlev\'{e} test, and the VSPE~\eqref{my2} can be strongly expected to be integrable.

Let us remind, however, that the Painlev\'{e} property does not prove the integrability of a nonlinear equation but only gives a strong indication that the equation must be integrable. Consequently, the new probably integrable nonlinear systems~\eqref{my1} and~\eqref{my2}, discovered in this paper, deserve further investigation, especially taking into account their importance for physics and technology. To find their Lax representations, generalized symmetries, Hamiltonian structures and soliton solutions seems to be a complicated problem, but a sufficiently interesting one to attract attention of experts in nonlinear mathematical physics.

\bigskip

The main part of this research was carried out during the author's visit to the Max Planck Institute for Mathematics (Bonn, Germany), whose hospitality and support are acknowledged with a deep gratitude.

\end{document}